\documentclass[aps,pre,twocolumn,superscriptaddress,showpacs]{revtex4}
\usepackage{graphicx}

\newcommand{\bc}{{\bf c}}
\newcommand{\bP}{{\bf P}}

\begin{document}

\preprint{elastic\_anisotropy version 12 -- \today}

\title{
Modeling Dipolar and Quadrupolar Defect Structures generated by Chiral Islands in Freely-Suspended Liquid Crystal Films}

\author{N.~M.~Silvestre}
\email[]{nunos@cii.fc.ul.pt}
\affiliation{Departamento de F\'{\i}sica da Faculdade de Ci\^{e}ncias and}
\affiliation{Centro de F\'{\i}sica Te\'orica e Computacional,
Universidade de Lisboa,\\
Avenida Professor Gama Pinto 2, P-1649-003 Lisboa Codex, Portugal}
\author{ P.~Patr\'\i cio}
\affiliation{Instituto Superior de Engenharia de Lisboa\\
Rua Conselheiro Em\'\i dio Navarro 1, P-1949-014 Lisboa, Portugal}
\affiliation{Centro de F\'{\i}sica Te\'orica e Computacional,
Universidade de Lisboa,\\
Avenida Professor Gama Pinto 2, P-1649-003 Lisboa Codex, Portugal}
\author{ M.~M.~Telo da Gama}
\affiliation{Departamento de F\'{\i}sica da Faculdade de Ci\^{e}ncias and}
\affiliation{Centro de F\'{\i}sica Te\'orica e Computacional,
Universidade de Lisboa,\\
Avenida Professor Gama Pinto 2, P-1649-003 Lisboa Codex, Portugal}
\author{A.~Pattanaporkrattana}
\affiliation{Department of Physics and Liquid Crystal Materials Research Center, University of Colorado, Boulder, CO, USA}
\affiliation{Physics Department,
            Kasetsart University, Bangkok 10900, Thailand}
\author{C.~S.~Park}
\affiliation{Department of Physics and Liquid Crystal Materials Research Center, University of Colorado, Boulder, CO, USA}
\author{J.~E.~Maclennan}
\affiliation{Department of Physics and Liquid Crystal Materials Research Center, University of Colorado, Boulder, CO, USA}
\author{N.~A.~Clark}
\affiliation{Department of Physics and Liquid Crystal Materials Research Center, University of Colorado, Boulder, CO, USA}

\date{March 4, 2009}

\begin{abstract}
We report a detailed theoretical analysis of novel quadrupolar interactions
observed between islands, which are disk-like inclusions of extra
layers, floating in thin, freely suspended smectic C liquid crystal films.
Strong tangential anchoring at the island boundaries result in a strength $+1$~chiral defect in each island and a companion $-1$~defect in the film,
these forming a topological dipole. While islands of
the same handedness form linear chains with the topological dipoles pointing in the same
direction, as reported in the literature, islands with different handedness form compact
quadrupolar structures with the associated dipoles pointing in opposite
directions. The interaction between such heterochiral island--defect pairs
is complex, with the defects moving to minimize
the director field distortion as the distance between the islands changes.
The details of the inter-island potential and the trajectories of the $-1$~defects
depend strongly on the elastic anisotropy
of the liquid crystal, which can be modified in the experiments by varying the material chirality of the liquid crystal. A Landau model that describes the energetics of
freely mobile defects is solved numerically to find equilibrium configurations for
a wide range of parameters.
\end{abstract}


\maketitle

\section{Introduction \label{introduction}}

The interactions between colloidal particles, fluid drops, and phase-separated
domains in liquid crystals (LCs) are mediated by the distortion of the ordered
fluid and depend on the LC elastic constants as well as on the anchoring
conditions at the inclusion boundaries.
The tendency of such inclusions to form aggregates depends intimately on
the properties of the inclusions as well as those of the LC,
providing exquisite control of the self-assembled colloidal
structures. A key feature of these novel systems is the presence of
topological defects, which are singularities of the LC director field, close
to the boundaries of the inclusions. In the strong anchoring regime, the director
field is distorted from its equilibrium uniform state near each inclusion and, in general,
topological constraints lead to a distortion equivalent to that of a virtual
$+1$~defect inside the inclusion. In order to match the uniform director field far away,
one or more defects of opposite sign are nucleated in the liquid crystal
matrix, outside the inclusion. The nature of these defects determines the symmetry of
the director field around inclusions and their long-range
interactions, for example, whether they are dipolar or quadrupolar. The short-range
interactions are more difficult to predict as the defects may move when
the inclusions approach each other, leading to complex two-body interactions.

When small spherical particles are suspended in three-dimensional (3D) nematics
\cite{Stark.2001}, several beautiful colloid-defect structures are
observed \cite{Poulin.1997,Musevic.2006}, including the ``satellite''
(a single $-1$~defect next to an
inclusion with radial anchoring)
\cite{Ruhwandl.1997,Lubensky.1997,Stark.1999}, ``Saturn-ring''  (a defect-line
around an inclusion with radial anchoring) \cite{Gu.2000}, and ``boojum'' (a
pair of strength $1/2$ defects at the poles of inclusions with tangential
anchoring) configurations \cite{Poulin.1998}. The director field around the inclusion--defect pair
typically has dipolar or quadrupolar symmetry \cite{Lev.2002}, leading
to a variety of self-organized superstructures
\cite{Loudet.2000,Nazarenko.2001,Meeker.2000}.

In two dimensions (2D), the first experimental observation of inclusions of
nematic droplets in chiral smectic~C films reported radial anchoring of the director at the droplet boundaries and single
$-1$~defects nearby \cite{Cluzeau.2001}. Later observations of
nematic and isotropic inclusions
in achiral Sm~C smectic films reported tangential anchoring
\cite{Cluzeau.2002,Voltz.2004,Cluzeau.2005,Dolganov.2006}, and either one pair
of $-1/2$ defects or a single $-1$~defect at the inclusion boundaries. More
recently, a temperature-dependent tangential to radial anchoring transition at inclusion boundaries in chiral smectics with small spontaneous polarization was observed
\cite{Dolganov.2007,Fukuda.2007}, resulting in
positional rearrangements of the defects along the inclusion
boundaries.

In principle, 2D systems of inclusions are simpler to model numerically than 3D systems as the computations required to calculate inclusion interactions
coupled to mobile defects are considerably less demanding. Accurate simulations of a variety of 2D systems have
been reported over a wide range of physical parameters.
Important analytical developments based on techniques such as
series expansions \cite{Burylov.1994}, electromagnetic analogies
\cite{Pettey.1998}, and complex analysis \cite{Bohley.2006} have provided a wealth
of detailed information that complements the numerical analysis and have been used to
test the accuracy of the numerical results. The method of images was used
recently in this context to calculate the long-range interactions between
inclusions under a variety of conditions \cite{Korolev.2008} and
inclusion--defect pair configurations, at short and long range, have been
calculated numerically with very high precision
\cite{Fukuda.2001,Patricio.2002,Yamamoto.2004}. The models reveal
that the short-range interactions between quadrupolar inclusions are
complex, with the defects moving as the distance between the inclusions is changed.
When the islands are close together, the defects are shared equally by the two
inclusions at equilibrium \cite{Tasinkevych.2002}, reminiscent of covalent atomic
bonding where the electronic density peaks in the region between the atoms. (A
similar `bonding' mechanism where a circular defect line is shared by two inclusions, creating
some spectacular equilibrium configurations, was also predicted numerically and observed in 3D nematics \cite{Guzman.2003,Ravnik.2007}).

In this article, we study the interactions between circular islands,
which are disk-like inclusions of extra layers, in freely suspended Sm~C films \cite{Bohley.2008}. These
films are structured in 2D fluid layers, with the liquid crystal molecules aligned along an average direction that is tilted from the layer
normal. The projection of the molecular tilt on the plane of the layers defines
a 2D vector, the so-called $\bc$-director. The aim of this work is to provide a
quantitative description of recent observations of the interactions between topologically
chiral islands in both chiral and achiral freely suspended Sm~C films, in which the existence of
an important class of inclusion interactions not studied previously in either
2D or 3D, was revealed, namely interactions that depend on inclusion chirality
\cite{chart.expts}. The experiments showed that islands
often form linear chains with the dipoles aligned along the chains, similar to earlier observations, but that heterochiral island--defect pairs form a novel
quadrupolar structure with the topological dipoles pointing in opposite
directions. Indeed, both dipolar chains and quadrupolar
assemblies may coexist on the same achiral Sm~C film, the local organization being
determined by the topological chirality of the individual islands.

This paper is organized as follows. In Section II we briefly summarize
the experimental results, highlighting those that will be compared with the
numerical calculations.
In Section III we present a Landau free
energy model that describes self-consistently the disposition of freely mobile defects in the film, and solve
it numerically for a wide range of parameters in Section IV, where a detailed
comparison with the observations is carried out. Finally, we discuss the
results and summarize our conclusions in Section V.

\section{Experimental results}

\begin{figure}
\par\columnwidth=20.5pc
\hsize\columnwidth\global\linewidth\columnwidth
\displaywidth\columnwidth
\includegraphics[width=200pt]{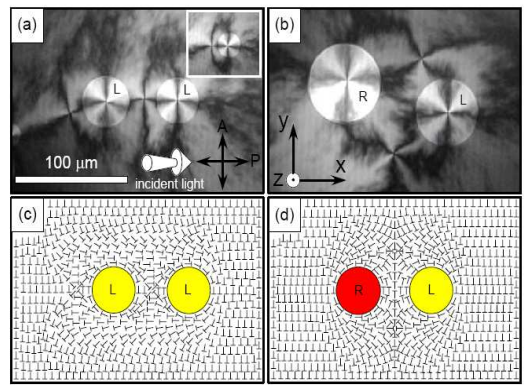}
\caption {Chiral islands in a freely-suspended smectic film of achiral liquid
crystal. DRLM images of a smectic C film of racemic MX8068 showing (a) two
islands containing $+1$~defects with the same handedness and (b) two islands
with $+1$~defects of opposite handedness. The defects in (a) form a dipolar
chain, while those in (b) form a topological quadrupole. The inset in (a) shows
a single island with a $+1$~defect inside and a matching $-1$~defect in the
background film, the pair forming a dipole. The laser illumination (shown by the arrow) is obliquely
incident in the x-z plane, allowing us to differentiate
between left- and right-handed islands by comparing the brightness of the
brushes in their upper and lower halves. The equilibrium $\bc$-director field
($\vdash$) is shown around two islands with (c) the same handedness and (d)
opposite handedness, the head of the $\vdash$ indicating the end of the molecule
closer to the viewer.} \label{fig1}
\end{figure}

Figure 1 illustrates characteristic equilibrium configurations of
island--defect pairs (with $N\sim 20$ layers) interacting on a background film
($N=2$) of racemic MX8068 \cite{mx8068}. Strong orientational anchoring at the island boundaries induces a
$+1$~defect inside each island and $-1$~defects in the background film, each
island--defect pair forming a topological dipole (inset, Fig.~1(a)). Such dipoles attract
over long distances through an elastic dipole-dipole interaction
and, when the islands are sufficiently close, reach an equilibrium separation
where the elastic distortion is minimized. Two types of island--defect pairs
are observed: either the pairs form linear chains with the dipoles pointing in
the same direction along the chain, as in Fig.~1(a), or the dipoles point in
opposite directions and the island--defect pairs form a quadrupolar structure,
with the $-1$~defects placed symmetrically above and below the line joining the
islands, as in Fig.~1(b). A sketch of the equilibrium $\bc$-director field when two
islands with the same handedness are close to each other is shown in Fig.~1(c),
while when the islands are heterochiral, the quadrupolar structure
shown in Fig.~1(d) is favorable \cite{chart.expts}.

The equilibrium center-to-center separation $D$ of islands of radius $R$ in dipolar chains was found experimentally to be close to $2 \sqrt{2} R$, the value predicted theoretically and confirmed in simulations using a free energy with a
single elastic constant \cite{Pettey.1998,Patricio.2002}. Islands in the
quadrupolar structure also exhibit a well-defined equilibrium
separation. When the island separation is altered using optical tweezers \cite{chart.tweezers},
the vertical separation $S$ of the $-1$~defects also changes. At sufficiently large island
separation, the quadrupolar symmetry is broken and the structure evolves
into two separate dipoles \cite{chart.expts}.

\begin{figure}
\par\columnwidth=20.5pc
\hsize\columnwidth\global\linewidth\columnwidth
\displaywidth\columnwidth
\includegraphics[width=180pt]{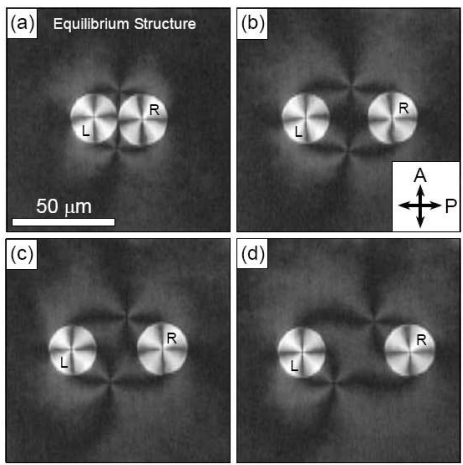}
\caption {Textures of heterochiral islands interacting on a film of 25\%
chirally doped MX8068. (a) The quadrupolar structure is in equilibrium when the
islands almost touch. (b) The equilibrium separation between $-1$~defects increases as
the islands are separated using optical tweezers. (c) When the
separation is sufficiently large, the quadrupolar symmetry is broken. (d) When the islands are
forced even further apart, the quadrupole evolves into two separate dipoles.}
\label{fig2}
\end{figure}

The interaction between islands on an $N=2$ achiral Sm~C film was found to be
weak, with large orientational fluctuations of the
$\bc$-director leading to large fluctuations in the position of the $-1$~defect. On chiral MX8068 Sm~C* films, in contrast, director fluctuations are smaller and the interaction between islands is large enough to be measured using optical
tweezers. On such chiral films, all of the islands are
topologically homochiral and they form only dipolar chains. We propose that the
reason for this bias is that the spontaneous polarization of the chiral
Sm~C* phase prefers to point toward the liquid crystal--air interface at the
edges of the islands, inducing only left-handed $+1$~defects. In films of racemic MX8068 doped with a small amount of chiral
material, left-handed defects predominate but some right-handed ones are also
observed. At 25\% chiral fraction, for example, about 1\% of the observed defects are
right-handed. At this chiral fraction, the quadrupolar structure is found to be
in equilibrium when the islands nearly touch (see Figure 2(a)).

\begin{figure}
\par\columnwidth=20.5pc
\hsize\columnwidth\global\linewidth\columnwidth
\displaywidth\columnwidth
\includegraphics[width=220pt]{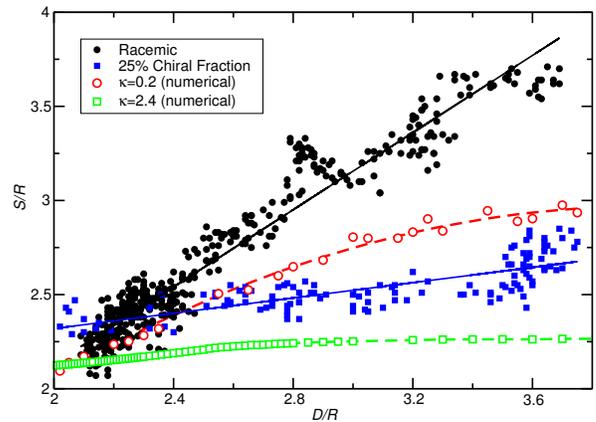}
\caption {
Equilibrium vertical separation $S$ of the $-1$~defects as a function of
the island center-to-center separation $D$ in the quadrupolar configuration
regime, for racemic and 25\% chirally doped films of MX8068, compared with the
results of numerical calculations for systems with elastic anisotropies
$\kappa=0.2$ and $\kappa=2.4$. The straight lines are linear fits of the
experimental data. 
The dashed lines are fits to the numerical data shown here to guide the eye. 
The value of $\kappa=2.4$ was chosen to reproduce the slope of the linear
fit for the chiral mixture. Note, however, that the theoretical results for
this $\kappa$ underestimate the value of $S/R$ at contact ($D/R=2$).
Fluctuations are
expected to increase the average defect separation, $S$, and account, at least
in part, for this discrepancy.
The average slope of the experimental data for the racemate is high, indicating very
small values of $\kappa$, a limit which is numerically inaccessible in our
simulations.
}
\label{fig3}
\end{figure}

Optical tweezers may be used to displace the islands from their equilibrium separations. As illustrated in Figs.~2(b)--(d), the equilibrium positions of the $-1$~defects change simultaneously as the islands are separated. In Figure 3 we show the experimentally measured vertical separation $S$ of the
$-1$~defects as a function of the island center-to-center separation $D$ for racemic and 25\%
chiral fraction films, plotted together with theoretical results to be described in the next section.

\section{Elastic free energy of Smectic~C layers}

The molecular orientation in smectic~C free standing films may be described by the $\bc$-director, a
two-dimensional unit vector field $\bc=(\cos\theta,\sin\theta)$ in the
direction of the in-plane projection of the average molecular tilt. The
total elastic free energy ${\cal F}$ of a film of thickness $\ell$ may be written
as the sum of two terms:
\begin{equation}
  {\cal F} = \ell\int_\Omega{f_{\bc}dxdy} + \ell\int_{\partial\Omega}{f_{_w}ds},
\label{tfe}
\end{equation}
where $\Omega$ is the area of the liquid crystal film, and $\partial\Omega$ its
boundary, including any inclusion boundaries that may be present. The second term accounts
for anchoring at the boundaries and may be omitted in the strong
anchoring regime. The first term describes the macroscopic free energy cost of
elastic distortions of the smectic~C director field and is given by
\begin{equation}
  f_{\bc}=\frac{1}{2}\left({\cal K}_s(\mbox{$\boldmath \nabla$}\cdot{\bf c})^2+{\cal K}_b(\mbox{$\boldmath \nabla$}\times{\bf c})^2\right)
\end{equation}
where ${\cal K}_s$ and ${\cal K}_b$ are respectively the 2D splay and bend elastic
constants.

This macroscopic elastic free energy neglects variations of the degree of liquid crystalline order
and it is inadequate when disordering effects are important. This is the
case when topological defects are present, as the order
vanishes at the center of the defects. A better approach, that is also more
tractable numerically, is to use a mesoscopic description that takes into
account variations of the degree of order on the scale of the relevant correlation
length. This is accomplished by theories of the Landau class
that consider the free energy as a functional of a generalized order
parameter, a quantity with the symmetry of the distortion field ($XY$ or 2D vector
in this case) that is non-zero in the ordered phase and vanishes
in the disordered phase. This order parameter varies throughout a non-uniform
system and is capable of describing the inhomogeneities in orientation and
degree of order that occur at topological defects.
If we replace the two-dimensional director field $\bc=(c_x,c_y)$ by the complex order parameter
$\phi=\phi_x+i\phi_y=|\phi|(c_x+ic_y)$, we may write the bulk free energy (in reduced
units \cite{Silvestre.2006}) as
\begin{equation}
  {\cal F}=\int_\Omega{f_{{\phi}}dxdy} \; {\rm ,}
\label{tfe_phi}
\end{equation}
with energy density
\begin{eqnarray}
  f_{\phi}&=&|\phi|^2(|\phi|^2-1)\nonumber\\
&+&\xi^2\left[(\partial_x\phi_x+\partial_y\phi_y)^2+\kappa(\partial_x\phi_y-\partial_y\phi_x)^2\right]
\; {\rm ,}
\label{eq21b}
\end{eqnarray}
where $\xi$ is the correlation length (the characteristic size of the inhomogeneities),
and $\kappa={\cal K}_b/{\cal K}_s$ the ratio of the elastic constants. This free energy includes all scalar invariants
of both the order parameter (up to fourth order in $|\phi|$) and its
derivatives. The negative sign of the quadratic term in $\phi$ indicates that the stable, uniform
phase is ordered and in the absence of constraints the free energy is minimized
by a uniform field of magnitude $|\phi|=1/\sqrt{2}$. When the direction of
$\phi$ varies in space, its magnitude may also change in order to minimize the total free energy.
At the defects, the order parameter
$\phi$ varies rapidly, its magnitude vanishing within a distance of the order
of the temperature-dependent correlation length $\xi$.

In chiral films, the director field is additionally coupled to the free energy associated
with the local space charge $-\nabla\cdot \bP$, where $\bP$ is the spontaneous
polarization, and is affected by screening from impurity ions dissolved in the film.
We assume here that the effect of material chirality of the LC may be
described by increasing the effective bend elastic constant, ${\cal K}_b$, or, equivalently,
by increasing the ratio of the elastic constants, $\kappa$ \cite{Meyer.1975,Lee.2007}.

\section{Simulation results}

In this section we present detailed numerical results describing the interaction
between homochiral and heterochiral islands for a range of parameters. In
particular, we calculate the separation of the two $-1$~defects associated with a pair of
heterochiral islands as a function of the island separation and
compare this with the experimental results to obtain an estimate of the
effective elastic anisotropy of the experimental systems.

We start by considering two islands with strong (fixed) tangential anchoring
conditions, and the same handedness. For
simplicity, we have taken $|\phi|=1/\sqrt{2}$, the bulk equilibrium value of
the order parameter for uniform alignment, at the island boundaries. At the borders of the domain,
representing large distances, we have forced the director to be uniformly
aligned along the $y$-direction. The correlation length that sets
the scale for the size of the defects was taken to be $\xi=0.1 R$, a tenth
the radius of the circular islands, $R$, which were assumed to be monodisperse.
This is admittedly an overestimate of the actual defect size but the major difficulty in the
numerical analysis stems from the disparity in the length scales $R$ and $\xi$, and
this choice is a compromise that speeds up the calculations and allows a systematic investigation of
a wide range of systems. As in previous work \cite{Patricio.2002}, we have used finite elements with
adaptive meshing to minimize the free energy.
A first triangulation respecting the predefined boundaries is
constructed. The order parameter $\phi$ is specified at the vertices of the mesh
and linearly interpolated within each triangle. The free energy is then
minimized using standard methods, the Hessian of the solution at each
iteration being used to generate a new mesh. If the solution presents stronger spatial
variations (usually near the defects), a mesh refinement is required.
On the contrary, if the solution
is almost constant over a large region (far from the islands, for example), a coarsening of the mesh is possible.
In typical calculations, convergence
is obtained after only two mesh adaptations, corresponding to final meshes with
$6\times10^3$ points, spanning a region of $15R\times15 R$, and minimal mesh
sizes of $10^{-4}R$ close to the defects. The free energy is
obtained with a relative accuracy of $10^{-4}$.

\begin{figure}
\par\columnwidth=20.5pc
\hsize\columnwidth\global\linewidth\columnwidth \displaywidth\columnwidth
\includegraphics[width=220pt]{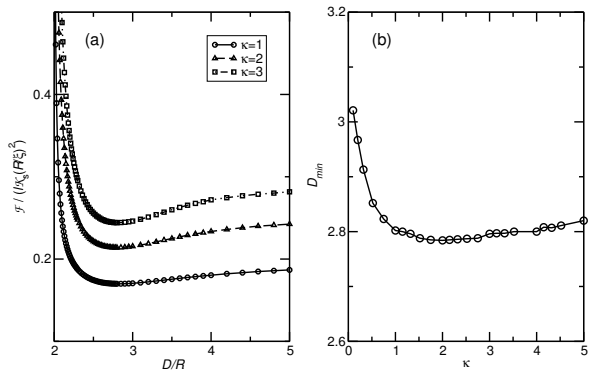}
\caption {(a) Total free energy as a function of homochiral island separation $D$,
for different $\kappa$.
When $\kappa={\cal K}_b/{\cal K}_s=1$, the free energy exhibits a well-defined
minimum at a separation $D_{\rm min}=2 \sqrt{2} R$. (b) Equilibrium separation
$D_{\rm min}$ as a function of the elastic anisotropy $\kappa$ of the system.}
\label{fig4}
\end{figure}

We have calculated the free energy of a pair of homochiral islands forming a chain with
their satellite $-1$~defects as a function of the island center-to-center separation $D$.
In order to investigate the effects
of the elastic anisotropy, $\kappa$ was varied over a wide range. We see
from Fig.~4(a) that the interaction between two
islands is attractive at long distances, exhibits a well defined minimum,
and is repulsive at small distances. For $\kappa=1$, the equilibrium
separation $D_{\rm min}=2.80R$
is slightly different from that obtained in our previous
work, $D_{\rm min}=2.82R$, based on the Frank elastic free
energy (\ref{tfe}) with no order parameter variation and pinned defects \cite{Patricio.2002}.
This difference is a consequence of the larger value used for $\xi$ here. Nevertheless, in order to carry out
extensive calculations and explore a reasonable range of parameters, we decided
to keep $\xi=0.1 R$, as the discrepancy in theoretical equilibrium distances obtained
is small compared with the uncertainty in the experimental data.
The equilibrium separation $D_{\rm min}$ as a
function of $\kappa$ is plotted in Fig.~4(b), showing the separation increasing sharply as the bend elasticity is reduced below ($\kappa \rightarrow 0$)
but approximately constant for $\kappa \agt 1$.

\begin{figure}
\par\columnwidth=20.5pc
\hsize\columnwidth\global\linewidth\columnwidth \displaywidth\columnwidth
\includegraphics[width=220pt]{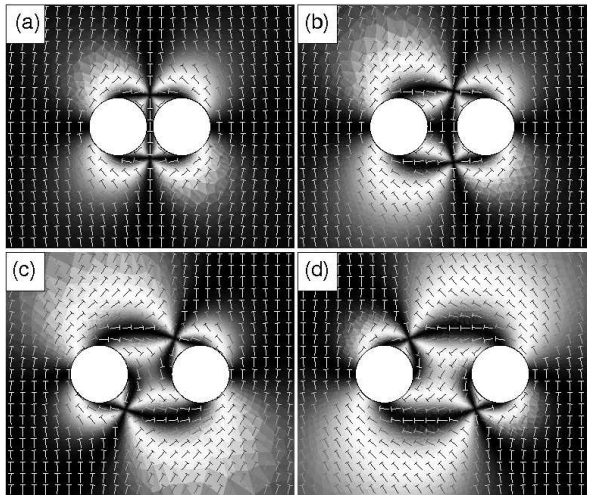}
\caption {Equilibrium director configurations for different heterochiral island separations with
$\kappa=1$. (a) $D=2.2R$: the defects are equally shared by the islands, in a symmetric,
quadrupolar configuration. (b) $D=3.0R$: the symmetry is
broken; here the defects are found in a metastable configuration where they are both closer to one island.  (c) $D=3.5R$: each defect follows a different island. (d) $D=4.0R$: even at this separation, the positions of the defects are still strongly influenced
by the presence of the other island--defect pair, with configurations typical of isolated dipoles occurring at
larger separations.}
\label{fig5}
\end{figure}

\begin{figure}
\par\columnwidth=20.5pc
\hsize\columnwidth\global\linewidth\columnwidth \displaywidth\columnwidth
\includegraphics[width=220pt]{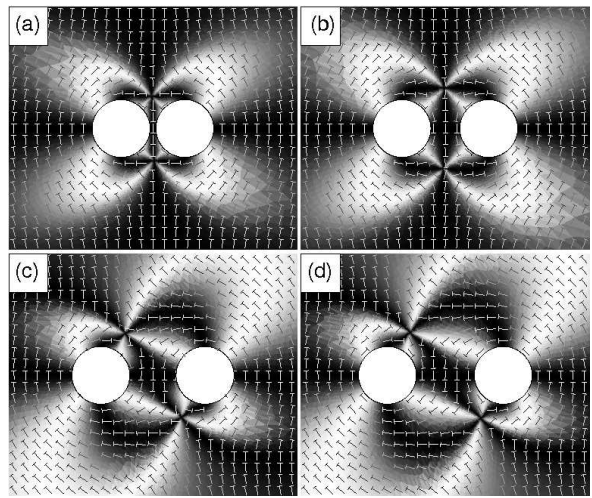}
\caption {Equilibrium director configurations for different heterochiral island separations with
$\kappa=0.2$. The order parameter around the $-1$~defects has no longer a circular symmetry as in Fig.~\ref{fig5}.
This is reflected in the narrowness of the white brushes that oome out of the $-1$~defects.
(a) $D=2.2R$. (b) $D=3.0R$. (c) $D=3.5R$. (d) $D=4.0R$.}
\label{fig6}
\end{figure}

We next considered two islands with opposite
handedness, with the defects located between them as observed in the
experiments and shown in Fig.~\ref{fig2}. In this case, the defects strongly influence the interaction
between the islands, and they move considerably as the distance between the islands
changes. As a result, the number of mesh adaptations had to be increased to
three or four in order to obtain reasonably accurate results (indicated by the
smoothness of the free energy plots, for example). Figures~\ref{fig5} and ~\ref{fig6} illustrate model
director configurations for different island separations $D$ for systems
characterized by two different elastic anisotropies, $\kappa=1$ and $\kappa=0.2$.
When the islands are very close together, the defects are located symmetrically in a quadrupolar
structure, being shared equally by the two islands (Figs.~\ref{fig5}(a), \ref{fig6}(a,b)). Above a certain island
separation, this symmetry is broken and each defect becomes associated primarily with just one of
the islands. We also sometimes observe  configurations in which
both defects are closer to the same island (see Fig.~\ref{fig5}(b)), although these are metastable. The defects
influence each other over large distances (Figs.~\ref{fig5}(c,d), \ref{fig6}(c,d)), their relative
positions with respect to their parent islands being very different from an
isolated island--defect configuration. When the elastic anisotropy of the liquid
crystal is large, the defect brushes become sharper  \cite{link.2005},
as shown in Fig.~~\ref{fig6} for the case ${\cal K}_s=5{\cal K}_b$
($\kappa=0.2$). The interaction between the islands is,
however, qualitatively similar to the isotropic elasticity case.

\begin{figure}
\par\columnwidth=20.5pc
\hsize\columnwidth\global\linewidth\columnwidth \displaywidth\columnwidth
\includegraphics[width=220pt]{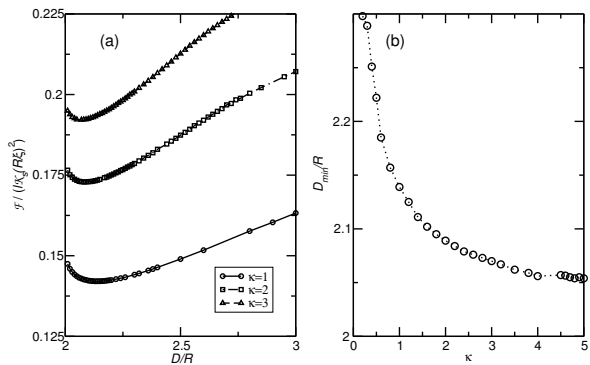}
\caption {
(a) Total free energy as a function of heterochiral island separation $D$,
for $\kappa={\cal K}_b/{\cal K}_s=1$, $2$, and $3$. All systems have a well defined
equilibrium separation $D_{\rm min}$ that is smaller than in the homochiral case and decreases monotonically with increasing $\kappa$  (b).}
\label{fig7}
\end{figure}

The effect of elastic anisotropy on the interaction between two heterochiral
islands was investigated by calculating the free energy as a function of the
center-to-center island separation, $D$, for different values of $\kappa$.  The results are
plotted in Fig.~~\ref{fig7}(a). For small $\kappa$, the potential is similar to that
observed with homochiral islands (Fig.~~\ref{fig4}(a)) but the equilibrium
separation is smaller and decreases monotonically with $\kappa$, as
shown in Fig.~~\ref{fig7}(b). The islands are much closer together than in the homochiral case and when $\kappa$ is large, the islands almost
touch each other, as observed in the experiments \cite{chart.expts}.

We next studied the geometric evolution of heterochiral island--defect pairs as
the island separation was varied. In particular, we tracked the vertical separation
of the $-1$~defects as a function of the distance between islands in order to
compare it with the experimental results and to estimate the effective elastic
anisotropy of the experimental systems.

\begin{figure}
\par\columnwidth=20.5pc
\hsize\columnwidth\global\linewidth\columnwidth \displaywidth\columnwidth
\includegraphics[width=220pt]{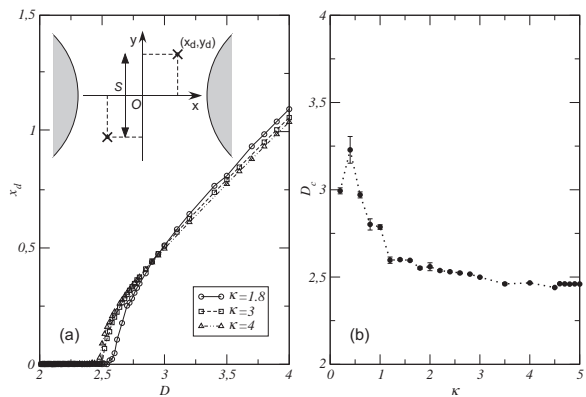}
\caption {(a) Computed horizontal position of the right defect $x_d$ as a function of
island separation $D$, for $\kappa=1.8$, $3.0$, and $4.0$. The reference system
has its origin at the mid-point between the islands. (b) The quadrupolar symmetry of the
island--defect pair is generally broken at a critical island separation $D_c$,  shown here as a function of $\kappa$.
The noisiness of the data when $\kappa < 1$ reflects the numerical difficulties
encountered in modeling this elasticity regime.}
\label{fig8}
\end{figure}

Determining the trajectories of the $-1$~defects as a function of island
separation is computationally very demanding, as an accurate determination of
the position of the defects requires a large number of mesh adaptations, and is
particularly difficult at the quadrupolar symmetry breaking transition, an effect we now describe. At
small separations, the islands share the defects in a symmetric, quadrupolar
configuration. In this regime, the defects are along the vertical bisector of
the islands, so their horizontal coordinate $x_d = 0$ (see Fig.~\ref{fig8}(a))
but at a critical island separation, $D_c$, the quadrupolar symmetry is broken
and $x_d$ becomes non-zero. One of the defects moves to the right, acquiring a
positive $x_d$ offset, while the other moves symmetrically to the left and has
negative $x_d$. This behavior is qualitatively the same for all values of the
elastic anisotropy $\kappa$ that were investigated, but the island separation
at which the transition occurs depends on $\kappa$. Fig.~\ref{fig8}(b) shows
the critical island separation $D_c$ as a function of $\kappa$. The trend
observed in $D_c$ confirms that the quadrupolar symmetry is broken in a
continuous transition, or bifurcation, that occurs at a precise value of the
island separation.

\begin{figure}
\par\columnwidth=20.5pc
\hsize\columnwidth\global\linewidth\columnwidth \displaywidth\columnwidth
\includegraphics[width=220pt]{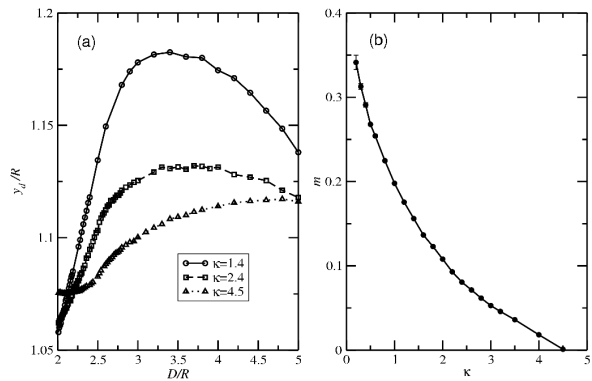}
\caption {(a) Computed vertical displacement $y_d$ of the upper $-1$~defect as a
function of island separation $D$, for $\kappa=1.4$, $2.4$, and $4.5$ [$y_d$ is
half the vertical defect separation $S$ plotted in Fig.~\ref{fig3}.] For small $D$
(close to equilibrium), the displacement of the defect increases linearly with
a well-defined slope $m$, shown in (b) as a function of the elastic anisotropy
$\kappa$.} \label{fig9}
\end{figure}

Similarly, the vertical displacement of the defects $y_d$  was determined as a
function of $D$, with results shown in Fig.~\ref{fig9}(a) for several values of
$\kappa$. For moderately anisotropic systems ($\kappa=1.4$ and $2.4$), the vertical displacement of the defects first increases, approximately linearly, as
the island separation increases. When $x_d$ bifurcates at $D_c \approx 2.5
R$, however, the behavior of $y_d$ changes, increasing to a maximum and
then decreasing slowly, until it reaches the isolated island--defect value,
$y_d=0$, at large separation. These results illustrate how strongly the
island--defect pairs interact even at quite large distances: for $D\sim 5 R$,
the vertical displacement of the defects is still very large, $y_d\sim R$. The
initial slope of the $y_d$ vs. $D$ curve, which is a measure of the susceptibility of $y_d$ to fluctuations in the island separation, is plotted in Fig.~9(b) as a
function of $\kappa$. For values smaller than $\kappa=0.2$, the numerical
results become unreliable: although the
energy can still be obtained accurately in this regime, the equilibrium defect positions are sensitive to numerical noise and can not be uniquely defined.  As the islands separate,
the position of the defects may vary substantially, and due to the high degree
of accuracy required  for small $\kappa$, our method is not able to resolve
them.

The numerical vertical separation of the defects, $S=2y_d$, were compared with the experimental 
results in Fig.~\ref{fig3}.
The value of $\kappa=2.4$ was chosen to reproduce the slope of the linear
fit for the chiral mixture. Note, however, that the theoretical results for
this $\kappa$ underestimate the value of $S/R$ at contact ($D/R=2$).
Keeping the same $\kappa$, we decreased the correlation length $\xi$ by a factor of ten.
However, we still obtained a smaller value of $S/R$ at contact.
Fluctuations are
expected to increase the average defect separation, $S$, and account, at least
in part, for this discrepancy \cite{Korolev.2008}.
The average slope of the experimental data for the racemate is high, indicating very
small values of $\kappa$, a limit which is numerically inaccessible in our
simulations.

In principle, the susceptibility of the vertical displacement of the $-1$
defects to perturbations of the islands from their equilibrium separation
plotted in Fig.~\ref{fig9}(a) should allow an estimate of $\kappa$ in the
experiments. The calculations suggest a strong variation of this susceptibility
as the elastic anisotropy $\kappa$ changes, as shown in Fig.~\ref{fig9}(b), but
we have thus far been unable to reconcile the predicted behavior quantitatively
with the experimental data.

\section{Conclusion}

We have described numerical simulations of the
spontaneous quadrupolar symmetry breaking of the director field around
heterochiral island--defect pairs in Sm~C liquid crystal films. In these films,
the $\bc$-director is tangentially anchored at the island boundaries,
inducing a $+1$~defect inside each island and a $-1$~defect outside in the
background film, the island--defect pair forming a dipole. Such dipoles attract
each other over long distances through an elastic dipole-dipole interaction
and, when the islands are sufficiently close, reach an equilibrium separation
where the overall elastic distortion is minimized.

Two types of island--defect pairs observed in the experiments were considered
in the theoretical analysis. Either the islands have the same handedness, in
which case they form linear chains with the dipoles pointing in the same
direction along the chain, or the islands have different handedness and the
dipoles point in opposite directions. In the latter case, at short distances,
the island--defect pairs form a quadrupolar structure, with the $-1$~defects
placed symmetrically above and below the line joining the islands.

The simulations show that interactions between two heterochiral islands occur over a much longer range than homochiral island--defect pairs. Even
when the islands are separated by up to five times their radius, the
interaction between heterochiral island--defect pairs is still quite different
from the long range dipole-dipole interaction expected in the linear asymptotic
regime, and an island--defect configuration significantly different from that expected for isolated dipolar pairs.

We have observed, both experimentally and numerically, a transition where the
quadrupolar symmetry of the heterochiral island system is broken. This occurs at a well-defined critical island separation
that depends on the effective elastic anisotropy of the liquid crystal and the simulations
suggest that the quadrupolar symmetry is broken continuously. For each system,
characterized by an elastic anisotropy $\kappa$, a well-defined critical island
separation was found where the defects are displaced symmetrically from the
vertical mid-plane bisecting the islands. The island separation at which this
transition occurs decreases smoothly as $\kappa$ increases (at least for
$\kappa \agt 0.5$), giving further support to the existence of a bifurcation that delimits
the quadrupolar configuration regime. The numerical calculations of the defect
positions are computationally demanding, leading to somewhat noisy results
for systems with small $\kappa<1$.
Furthermore, at the transition large thermal fluctuations of the director
field are expected and were observed
experimentally in racemic systems. Including their effects in the model will
affect the location (and maybe even the
nature) of the bifurcation predicted by our mean-field analysis.

At small island separations, the computed vertical displacement of the
defects was found to increase linearly, in general agreement with experiment. The scatter in
the experimental data strongly suggests that thermal fluctuations in the
position of the defects are important even in the quadrupolar configuration regime and
that these fluctuations should be taken into account in a full description of
the symmetry breaking transition. The simulations suggest that the initial slope
of the $S$ vs.\ $D$ curve may be used to estimate the effective elastic
anisotropy $\kappa={\cal K}_b/{\cal K}_s$ of experimental systems. We have found,
however, that a good fit to the slope underestimates the defect separation
at contact. It is likely that thermal fluctuations, which increase the
average separation between defects, are responsible for this discrepancy.
Furthermore, the polydispersity of the islands may also affect
the simulation results reported here. Heterochiral islands of
different sizes do not exhibit perfect quadrupolar symmetry and this may
result in a different behaviour of the experimental defect separation. Islands with
different sizes and the effect of thermal fluctuations will be investigated in
future work.

\begin{acknowledgments}
Financial support from the Foundation of the University
of Lisbon and the Portuguese Foundation for Science and Technology
FCT under Contracts Nos. POCI/FIS/58140/2004 and POCTI/ISFL/2/618 is gratefully acknowledged.
NMS acknowledges the support of FCT through grant No. SFRH/BPD/40327/2007.
The Colorado group was supported by NASA Grant NAG-NNC04GA50G and by NSF MRSEC Grant No.~DMR~0213918.
\end{acknowledgments}

\end{document}